\documentstyle[aps,prl,multicol,psfig]{revtex}
\input epsf

\newcommand{\beq}{\begin{equation}}
\newcommand{\eeq}{\end{equation}}
\newcommand{\bea}{\begin{eqnarray}}
\newcommand{\eea}{\end{eqnarray}}
\def\lsim{\mathrel{\rlap{\lower4pt\hbox{\hskip1pt$\sim$}}
    \raise1pt\hbox{$<$}}}         %less than or approx. symbol
\def\gsim{\mathrel{\rlap{\lower4pt\hbox{\hskip1pt$\sim$}}
    \raise1pt\hbox{$>$}}}         %greater than or approx. symbol

\def\bqn{\begin{eqnarray*}}
\def\eqn{\end{eqnarray*}}

\begin{document}

\draft

\title{\bf Lattice model for cold and warm swelling of polymers in water}

\author{Pierpaolo Bruscolini\cite{P} and Lapo Casetti\cite{lapo}}

\address{Istituto Nazionale per la Fisica della Materia (INFM) and
Dipartimento di Fisica, Politecnico di Torino, \\
Corso Duca degli Abruzzi 24, I-10129 Torino, Italy} 

\date{\today}

\maketitle

\begin{abstract}
We define a lattice model for the interaction of a polymer with water.
We solve the model in a suitable approximation. In the case of a 
non-polar homopolymer, 
for reasonable values of the parameters, the polymer is found
in a non-compact conformation at low temperature; as the  temperature grows, there is a sharp transition towards a 
compact state, then, at higher temperatures, the polymer swells again.  
This behaviour  closely reminds that  of proteins, that are unfolded  
at both  low and high temperatures.
\end{abstract}

\begin{multicols}{2}
\narrowtext

The theoretical modelling of the behaviour of polymers in aqueous
solution is a long-standing problem which has received considerable
attention also in connection with the problem of protein folding.  It
is commonly believed \cite{Creighton} that the organization of water
molecules in quasi-ordered structures around non-polar monomers plays
a fundamental role in stabilizing the folded ``native'' state of a protein, 
namely, the biologically active state.  In simple models of protein
folding, however, water is usually taken into account within the
definition of the interactions between aminoacids, rather than
explicitly considering water-monomer interactions.  The rationale for
this approach \cite{GaOra96} is that water behaves as a bad solvent
for hydrophobic monomers, and this can be modeled by an effective
Hamiltonian with attractive monomer-monomer interactions. However, an
``exact'' effective Hamiltonian, obtained from a partial trace over
water degrees of freedom, should in principle depend also on the
temperature and on the conformation of the polymer.  Hence it is
interesting to study simple models where the role of water can be
considered explicitly.

Here we introduce a lattice model where each site is occupied either
by a monomer or by water molecules.  
Water-monomer interactions depend
on the state of water; water-water interactions are dealt with in a
coarse-grained way. To distinguish between the effect of the presence 
of water and any
other, monomer-monomer interactions  
are neglected. 
The model can be thought of as
a  generalization of the one proposed by De Los Rios and Caldarelli (DLRC)
\cite{coldunfo}. Our generalization is an attempt to model
more carefully the water degrees of freedom, yet keeping the description
as simple as to obtain a model which can be approached by analytical 
techniques. 
Specializing to the case of a non-polar homopolymer, we introduce  
a suitable approximation scheme and
evaluate analytically the partition function, the specific heat and
the average number of water-polymer contacts, which is a measure of
the compactness of the polymer conformation.  
We find that, for reasonable values of the
parameters, the polymer behaves as follows: at low temperatures, it is
found in a non-compact conformation. As the temperature grows, there
is a sharp transition towards a compact state, associated with a sharp
peak in the specific heat of the water-polymer system; then, at higher
temperatures, the polymer swells again, smoothly.  This behaviour
reminds that of proteins, that are unfolded at both low and
high temperatures \cite{Privalov}.

Let us now define our model.
We consider a $d$-dimensional lattice with $N_s$ sites and
coordination number $z$. Each lattice site is occupied either by a
monomer of the polymer or by a water ``cluster''.  The polymer
is made of $N$ monomers. Each monomer can be either hydrophobic (H) or polar (P), to which cases we associate 
$\sigma = 0,1$, respectively. The sequence of H's and P's is fixed, so that
the $\sigma$'s are not dynamical variables.  
The number of water
clusters is $N_w = N_s - N$. Each water cluster is a
system of $m$ water  molecules, to account for the fact that several
water molecules are affected by the presence of a monomer.
For each cluster, the state-space has an ordinary part,
characterized by $\chi = 0$ and made up of an infinity of states $\vartheta =
0,1,2,\ldots,\infty$,
and a special part, with  $\chi =1$, containing only the state 
``*''.
This state is special in the sense that it more favourably interacts
with hydrophobic monomers, as we will see in the following. The choices
of a discrete spectrum and of a unique special state are made
for the sake of simplicity. 
 
The model Hamiltonian is  written as a sum of a non-interacting water
term plus a water-polymer  interaction; no energy is associated to  the
polymer alone:
\beq
{\cal H} ={\cal H}_{\text{water}}+{\cal H}_{\text{int}}~.
\label{eq:h=h0+hint}
\eeq
The first term has the form:
\beq 
{\cal H}_{\text{water}}= \sum_{j = 1}^{N_w} \left[ \vartheta_j \left(
1-\chi_j \right) + E^* \, \chi_j  \right]~,
\label{eq:h0}
\eeq
where $E^* >0$.
For each label $j$ the state $\vartheta_j = 0$ represents ice, i.e., a 
completely ordered
state, with two hydrogen bonds per molecule.  Each water-filled
site, representing  a cell of $m$ molecules, 
in the $\vartheta = 0$ state has $2 \, m$ hydrogen-bonds, a fraction of which 
is buried in the bulk of  the cell, while the rest connects
neighboring cells; we set the energy of this state equal to zero.
The states $(\chi_j=0;~\vartheta_j = 1,2, \ldots)$ are excited states obtained 
by adding kinetic energy to the water cluster: the hydrogen bonds are weakened
and, eventually, broken. 
According to this coarse-grained description, which disregards the
details of water-water interactions, the degeneracy $g(\vartheta)$ 
of the energy level
$\vartheta$ is given by the number of ways in which one can assign
$\vartheta$ ``quanta'' of energy to the $m$ molecules, i.e., 
the number of ordered
partitions of $\vartheta$ objects in $m$ classes: 
$g(\vartheta) = 
{ m + \vartheta - 1 \choose \vartheta}$.

The special state $(\chi=1; ~*)$ models 
the partial ordering that water molecules are believed to assume in
presence of a non-polar residue \cite{Creighton}. 
It is indeed an excited states of energy $E^*$, with a 
peculiar character: the $m$ water molecules are thought of as 
geometrically ordered in the cell, 
so that hydrogen-bond weakening and
breaking affects only the surface molecules in the water-filled site.
An accurate evaluation of the degeneracy  $g^*$ of this state would require a
detailed specification of the microscopic geometry of water, which is
beyond the scope of the present analysis; nevertheless  a
crude estimate of it can be given 
by noticing that, in the presence of a
water-monomer contact, 
$m_{\text{eff}} \approx z m^{\frac{d-1}{d}}$ water molecules shall
rearrange in an ordered conformation to maximize hydrogen-bonding.
Then, reasoning as before,
$g^*$ will be given by the number of ways to assign
$\vartheta^*=E^*$ quanta to the  $m_{\text{eff}}$ molecules, i.e.,
$g^*={[[m_{\text{eff}}]]+\vartheta^*-1 \choose \vartheta^*}$, 
where $[[\bullet]]$ is the integer part of $\bullet$. 

The second term in the Hamiltonian
(\ref{eq:h=h0+hint}) is
\beq
{\cal H}_{\text{int}} =  \sum_{i = 1}^N \Delta_{ij} \, \sigma_i 
\left[ J\, \chi_j  + K \left( 1 -
\chi_j \right) 
\right]~,
\label{eq:hint}
\eeq 
where $\Delta_{ij} = 1$ if $i$ and $j$ are nearest-neighbors and
$\Delta_{ij} = 0$ otherwise, $K>0$ (resp.\ $J \le 0$) is the energy
cost (resp. gain) of a
contact between a H monomer and a water site 
in an ordinary state (resp. special *-state). 
The form of Eq.\ (\ref{eq:hint}) can be understood if we consider a 
``droplet''  of  hydrophobic monomers in a water-filled
lattice, and analyze the energy balance in exchanging a water site with
a monomer  (both taken from the respective  bulks). 
If the water is in an ordinary
state, each contact with a  monomer will break up to 
$\sim
m^{\frac{d-1}{d}}$  hydrogen bonds between neighbouring water cells. 
If $a>0$ is the energy involved in a bond (in units of
the spacing between the levels), 
this process yields an energy cost of 
$K \simeq m^{\frac{d-1}{d}} a$, 
while the cost will be  zero if the water is in  the special *-state.
At the same time, non-polar residues interact more
favourably with water
than with each other, due to the permanent dipole moment of water 
\cite{dipole}.
This is taken into account by assuming an energy gain $J$ ($- a 
\lsim J \le 0$) for a contact between water in the *-state and a 
hydrophobic monomer.
It should be clear that this theoretical
framework is not meant  to provide  a  detailed microscopic
description of the physics of water, but rather to account for the
basic ingredients of non-polar solvation: the existence of an ice-like
ground state of zero entropy for water, and of a set of excited states,
some of  which may be particularly suited for interaction with non polar
solutes, but involve a substantial entropy loss.

Let us now study the equilibrium thermodynamical properties of the
model. To compute the partition function it is useful to
introduce some notations.  The maximum number of contacts between
water and H monomers is obtained when the polymer is in an extended
configuration, so that $M = (z - 2) N_H + \sigma_1 + \sigma_N$, where
$N_H$ is the number of non-polar residues.  
Given a
configuration $C$ of the polymer, the energy of the system can be
written as
\beq
{\cal H}(C) = \sum_{j = 1}^{N_w} \left[ \left(1-\chi_j \right) 
\left(\vartheta_j  + K \ell_j \right) +
\chi_j \left( E^* + J \ell_j \right)
\right]~,
\label{H(C)}
\eeq
where $\ell_j \equiv \ell_j(C)$ is the number of contacts between the 
$j$-th water  cluster and H monomers. 
Denoting by $n_\ell (C)$ 
the number of water clusters that have $\ell$ contacts 
with $H$ monomers, 
the partition function of the model can be written as 
\begin{mathletters}
\bea
&& {\cal Z} = \sum_C {\cal Z}(C)= \sum_{\{n_\ell\}} \zeta(\{n_\ell (C)\};N)\, 
Z (\{n_\ell (C)\})~; \\
&& ~~~~~~~~~~~~~Z (\{n_\ell (C)\}) = 
\sum_{\{\vartheta\}} e^{-\beta {\cal H}(C)}~,
\eea
\label{calZ}
\end{mathletters}
\noindent 
where $\zeta(\{n_\ell (C)\};N)$ is the number of equivalent 
conformations for a  polymer of length N, i.e., of the conformations with 
the same set of $n_\ell$'s. Now, to ease the notation 
let us put $Z(C) = Z (\{n_\ell (C)\})$. 
Using Eq.\ (\ref{H(C)}), and the fact that 
water cells do not interact with each other, so that it is possible to
factor $Z(C)$ according  to the labels  $j=1, \ldots, N_w$,  
$Z(C)$ can be written as
\begin{mathletters}
\beq
Z (C) = P_0^{N_w} \prod_{\ell = 1}^z x_\ell^{n_\ell (C)}~,
\label{Z_inf_1}
\eeq
where $x_\ell={P_\ell}/{P_0}$, and
\beq
P_\ell =g^*  e^{-\beta \left(E^*+ J \ell\right)} + e^{-\beta K \ell}
\frac{1}{\left( 1 - e^{-\beta}\right)^m}  ~.
\eeq
\label{Z_inf}
\end{mathletters}
\noindent
The mean energy and the specific heat are evaluated in the standard way as  
$U = \langle \cal{H} \rangle  = 
-\partial \log {\cal{Z}}/{\partial \beta}$ and $C_V = \partial U/\partial T$.
The average number of water sites presenting $\ell$
contacts with the polymer can be computed as 
$\langle n_\ell \rangle \;=\; x_\ell (\partial \log {\cal Z}/
{\partial x_\ell})$,
from which one can obtain the average (total) number of H-water
contacts, $\langle n_c \rangle = \sum_\ell \ell 
\langle n_\ell \rangle$,
which is a measure of the compactness of the polymer. 
Another interesting observable
is the average number of contacts with water sites in
the 
*-state, given by
$\langle n_c^*\rangle =- T \partial \log{\cal Z}/
{\partial J}$.

Notice that, even in the simple hydrophobic homopolymer case, our model can not be mapped onto a polymer model with monomer-monomer
contact interactions, because
 the partition function 
(\ref{calZ}) can not be written in the form
\beq
{\cal Z} = \sum_{n_{\text{HH}}} \xi(n_{\text{HH}})\, e^{-\beta h n_{\text{HH}}}
\label{ZHH}
\eeq
where $n_{\text{HH}} = (M - \sum \ell n_\ell)/2$ is
the number of monomer-monomer contacts and
 $\xi(n_{\text{HH}})$ is the number of polymer configurations with $n_{\text{HH}}$ internal contacts, with $h$ a true coupling constant, 
independent
of $T$ and $\{n_\ell\}$.

Before going further with the discussion of the thermodynamical
properties of our  model, let us 
note that the partition function of the DLRC model \cite{coldunfo}
can be obtained from the present one
with  the substitutions
$g^* \rightarrow 1$, $E^* \rightarrow 0$, $\left( 1 -
e^{-\beta}\right)^{-m} \rightarrow q-1$,
due to the fact that in the DLRC model all the states are equivalent
(with zero energy) for pure water, and when a water-monomer
interaction takes place,
the special state has zero entropy, while the 
degeneracy of the
excited states is independent of the temperature.
In both models one has to perform a sum over the
conformations of the polymer to calculate the thermodynamical functions.
This is a hard numerical task when the length of the
polymer grows: for this reason 
we introduce an approximation that allows us
to evaluate them analytically. First, 
we test our approximation against the exact
numerical results obtained by DLRC in the case of a homopolymer 
(with $N \leq 25$) 
on a 2-$d$ Manhattan lattice;  then we apply it to our model.

From now on 
we specialize to the case of a non-polar homopolymer
of length $N$ on a 2-d square lattice ($z=4$). Let us now come to the main point of our approach, i.e., let us  
introduce an approximation for 
the unknown quantity
$\zeta(\{ n_l(C) \};N)$ in Eq.\ (\ref{calZ}), representing the number of
polymer conformations characterized by the same set $\{n_\ell\}$.
We  observe that this number must be strongly dependent  
on $n_c$, the total number of water-monomer contacts, that in turn is related
to the number of monomer-monomer $n_{\text{HH}}$ contacts by $n_c= M - 2 n_{\text{HH}}$, 
and hence is a rough measure of how compact the
polymer is. Then, our approximation is built in the following three steps:
{\bf 1.} we assume  that $\zeta(\{ n_l(C) \};N)$ depends essentially on
$n_c$ and not on each of the $n_l$. More precisely, we assume 
that, at fixed number of contacts $n_c$, the number of
conformations in which there are water sites with three or more 
contacts with the
polymer is negligible (single and double contacts are much more
likely to appear, due to geometrical reasons), so that 
$\zeta(\{ n_\ell(C) \};N)$ does not depend on $n_3$ and $n_4$. Moreover, we assume also that $\zeta$ is the same for each $n_1$, $n_2$ such that $n_1 + 2 n_2=n_c$, i.e.,
\beq
\zeta(\{ n_\ell(C) \};N) = \zeta(n_c; N) \equiv C_N (n_{\text{HH}})  ~,
\eeq
where $C_N (n_{\text{HH}})$ is the
number of walks of length $N$ 
with $n_{\text{HH}}= (M-n_c)/2 $ internal contacts. 
{\bf 2.} We consider $\zeta( n_c; N )$ as  characterized by two
regimes,  a globular
and an extended one, referred to as $\zeta_c$ and $\zeta_g$, respectively,
which come into play depending on the value of $n_c$:
\begin{equation}
\zeta(\{ n_\ell(C) \})=  
\cases{ \zeta_{c}( n_c; N) &if $n_c> \tilde n_c$ ;
        \cr
        \zeta_{g}( n_c ; N) &if $n_c \le \tilde n_c$ .}
\label{eq:gradino}
\end{equation}
This is justified by the fact that the polymer shows, at different
temperatures, either a globular, compact phase or an extended one. The
latter involves a high number of water-monomer contacts, $n_c$, 
while the former is characterized by small values of $n_c$,
namely, $n_c \propto N^\frac{d-1}{d}$, as in the case of a maximally compact
state. We assume that a step function in the number of contacts
separates these regimes; the position of the step, $\tilde n_c$, is 
at present unknown.
{\bf 3.} We consider the extended conformations as self avoiding walks (SAWs)
and the
compact ones as Hamiltonian walks (HWs): their numbers are both exponential
in $N$, and are related by 
\begin{mathletters}
\beq
\zeta_{c}(N) \simeq \exp(\alpha N) \zeta_{g}(N)~,
\label{eq:number_of_walks}
\eeq
where $\alpha$ is given by
\beq
\alpha=\log \mu_{\text{SAW}} - \log \mu_{\text{HW}}
\label{eq:alpha}
\eeq
\label{SAW-HW}
\end{mathletters}
and $\mu_{\text{SAW}}$, $\mu_{\text{HW}}$ 
are the respective connectivity constants \cite{polymers}.

Resorting to the above assumptions we can evaluate the partition
function. In the homopolymer case $n_c$ is always even, so that we
set $k=n_c/2$, $k^{\text{min}}=n_c^{\text{min}}/2$, where 
$n_c^{\text{min}} \simeq z\sqrt{N}$ 
is the minimum number of contacts, $\tilde k=\tilde n_c/2$  
and sum over $k$, obtaining 
\begin{mathletters}
\bea 
{\cal Z} & = & P_0^{N_w} \zeta_{g}(N)
\left[ \psi(x_1,x_2;k^{\text{min}},\tilde k) \right. \nonumber \\
& + & \left. e^{\alpha N} \psi(x_1,x_2;\tilde k + 1, M/2)  \right]~, 
\eea
where $\psi$ is given by
\bea
&& \psi(x,y;p,q) =  \frac{x^{2p}}{(x^2 -1)(x^2 - y)(y-1)}\times \nonumber \\  
&& \left\{ x^2 + x^{2(q-p+2)} (y-1) 
  - y \left[ 1 + (x^2 - 1)y^{q-p+1} \right] \right\}.
\eea
\label{eq:zetapprox}
\end{mathletters}
The thermodynamical observables, $U$, $C_V$, $\langle n_\ell \rangle$ and 
$\langle n^*_c \rangle$
are then computed according to their definitions.

The approximation we introduced contains the threshold  parameter,
$\tilde n_c$, which is not a free parameter, but is unknown;
its evaluation  would
require the knowledge of the number of walks 
of arbitrary  length and number of internal contacts on a lattice.
In order to circumvent this difficulty, 
we fix $\tilde n_c$ applying our approximation to the DLRC
model, for which we know the results from exact enumeration on a 2-$d$ 
Manhattan lattice \cite {coldunfo}. Using our method, the best approximation 
of the DLRC model specific heat is obtained using with  
$\tilde n_c= n_c^{\text{min}} \, 23/20$; we choose this value 
for $\tilde n_c$.
For the parameter $\alpha$, Eq.~(\ref{eq:alpha}) holds, where, for a
Manhattan lattice, $\mu_{\text{SAW}} = 1.7335$ \cite{BeCaa98} and 
$\mu_{\text{HW}} 
\approx \exp(G/\pi)=1.3385$ ($G$ is
the Catalan constant) \cite{DuDa88}.

\begin{figure}
\epsfysize= 5 truecm 
\begin{center}
\mbox{\epsfbox{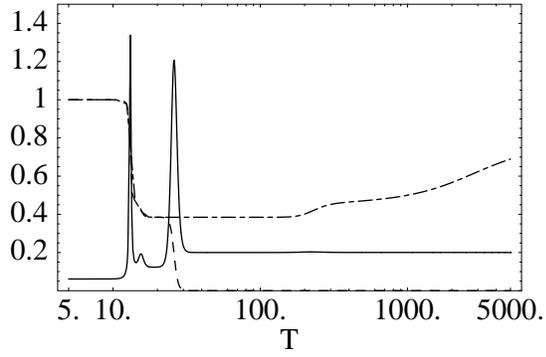}} 
\end{center}
\caption
{Specific heat $c = C_V/(m M)$ (continuous line, rescaled by 
a factor of 0.2 for graphical reasons), average number of 
water-monomer contacts $\langle n_c/M \rangle $ (dash-dotted), and average number of contacts with water in the *-state
$\langle n_c^{*}/M \rangle $ (dashed) as a function of
$T$. Here $N = 25$ and  $J= - E^*/2$. 
}
\label{fig:c_n_n1_n2}
\end{figure}

\begin{figure}
\epsfysize= 5 truecm 
\begin{center}
\mbox{\epsfbox{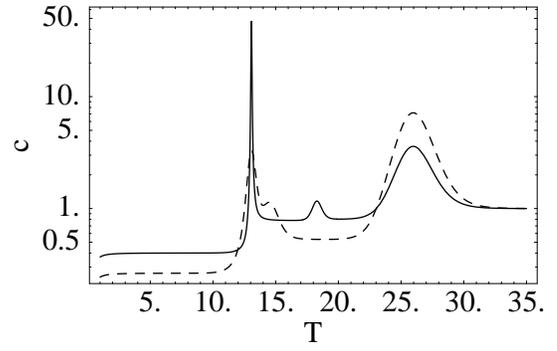}} 
\end{center}
\caption{$c = C_V/(m M)$ 
for $N=16$
(dashed) and $N=100$ (continuous).
$J= - E^*/2$.}
\label{fig:many_c}
\end{figure}

Let us now move to our model. Since the discretization of the water energy levels is artificial, the ratio $a$ of the hydrogen bond energy to the level
spacing is a free parameter. We choose $a = 100$; yet, we verified that even remarkable  changes in $a$ do
not imply major modifications of the behaviour of the thermodynamic quantities. 
The values of the other parameters have been fixed according to their physical meaning, i.e., we have chosen $m = 20$, $E^* = 2a$, $K = a \sqrt{m}$. Given the length $N$ of the polymer,
we set $N_w = M = 2N + 2$ (which is the maximum number of
sites which can wet the chain).
If $J \lsim - E^*/2$,   
at low $T$ 
the polymer is extended and completely wet by
the solvent, as should be a neighbor-avoiding walk.
Then, raising the temperature above a certain $T=T_f \simeq 13$, 
the polymer becomes compact ($\langle n_c\rangle $
drops to that of a maximally compact conformation), and finally, as $T$ grows 
further, it swells again   smoothly 
 (see Fig.~\ref{fig:c_n_n1_n2}).
The specific heat 
has a peak at $T_f$ and another one at a  higher
temperature, when the polymer is compact. The latter is
related to the excitation of the water sites
around the polymer in compact conformation out of the *-state, as witnessed
by the drop of $\langle n^*_c \rangle$ to zero in correspondence of the peak; the former peak at $T_f$, 
whose height grows with the length $N$ (see Fig.\ \ref{fig:many_c}), could be related to a true phase transition, reminding cold
unfolding in proteins.  The thermal swelling
 of the polymer is present here, but does not have the characteristic of a 
phase transition, as it happens, for instance, in \cite{HaJea99}. 
Yet  this is  not a surprise, due to the  
absence of any kind of imposed
cooperativity in our model. 
It is interesting to notice that, at low temperatures, the polymer is
expanded even for moderate values of $J$, i.e., $J  \approx -E^*/2$: 
this means that, in the presence of a monomer, water prefers to stay in
the excited special  state, even when the
energetic gain $J$ does not apparently compensate the energy loss
$E^*$. 
For values of $J$ closer to zero 
the cold collapse is lost: at low temperatures the polymer is compact, then it
swells smoothly  at high temperatures. 

To summarize, we have defined and discussed a lattice model for water-polymer 
interaction, where water is explicitly considered, 
and  introduced an
approximation scheme that allowed us to  analytically evaluate the 
relevant thermodynamical  averages. 
We remark that our model can not be mapped onto a model with monomer-monomer
contact interactions alone.
We addressed the homopolymer case and  
observed that, for reasonable values of the parameters, the polymer is
in a compact conformation at intermediate temperatures,  and
swells when the temperature is lowered as well as when it is raised. 
This
recalls similar proteins' behaviour.
While the ``cold unfolding'' process is sharp, and possibly represents a phase
transition as $N \to \infty$,  thermal swelling is smooth. 
The absence of a sharp transition at high temperature could possibly
be attributed to the lack of specificity in interactions: we expect that
the heteropolymer case (to be studied next) will show more  
similarities 
to real proteins.
The fact that our results are indeed qualitatively similar to those of 
DLRC \cite{coldunfo}, 
in spite of the differences in the model studied (many states 
{\em vs.} two, temperature-dependent degeneracy, etc.), suggests that the 
polymer behaviour we both find out is not a peculiarity of a particular model, 
but a property of a class of them: it is likely that the crucial thing 
is to take into account explicitly the degrees of freedom of the solvent, 
though in a simplified way. Indeed it appears 
(see e.g.\ Ref.\ \cite{Trovato}, where
a model of random heteropolymers is introduced and studied) that if the 
solvent's degrees are neglected from the beginning, and hydrophobicity 
is attributed to monomers like a ``charge'', cold unfolding will not 
be found for non-polar homopolymers, and probably even for quenched 
random heteropolymers.

We thank P.\ De Los Rios for useful discussions and criticism and for having provided us access to his results prior to publication, and M.\ Rasetti for a
critical reading of the manuscript. 
We also acknowledge A.\ Pelizzola, F.\ Seno and G.\ Tiana for fruitful discussions.

\end{multicols}
\end{document}